\documentclass[twocolumn]{article}
\pdfoutput=1
                        \newif\ifpaper \newif\ifPDF               
                        \newif\ifOUP \newif\ifboyscout            
                        \newif\ifdasbuch \newif\ifarticle         
                        \newif\ifsolutions                        
                        \dasbuchtrue 
                        \boyscouttrue 
                        \solutionstrue 
                        \paperfalse\PDFtrue 
                        \OUPfalse \articlefalse  

\usepackage{nolta2013}
\usepackage{txfonts}
\usepackage[latin1]{inputenc}
\usepackage[T1]{fontenc}
\usepackage{times}
\usepackage[pdftex]{graphicx}
\usepackage{array}
\usepackage[pdftex,colorlinks]{hyperref}

\ifpaper 
       \newcommand{\href}[2]{{#2}}  
       
       \renewcommand{\color}[1]{}       

\else 

\fi

\newsavebox{\bartName}

{\hspace*{\fill}\nolinebreak[1]\usebox{\bartName}\vspace*{1ex}\end{minipage}}
\newcommand{\beq}{\begin{equation}}
\newcommand{\continue}{\nonumber \\ }

\newcommand{\eeq}{\end{equation}}
\newcommand{\ee}[1] {\label{#1} \end{equation}}
\newcommand{\bea}{\begin{eqnarray}}

\newcommand{\eea}{\end{eqnarray}}
\newcommand{\barr}{\begin{array}}
\newcommand{\earr}{\end{array}}

\newcommand{\rf}     [1] {~\cite{#1}}

\newcommand{\refeq}  [1] {(\ref{#1})}

\newcommand{\reffig} [1] {figure~\ref{#1}}

\newcommand{\refsect}[1] {sect.~\ref{#1}}

\newcommand{\ie}{{i.e.}}        

\newcommand{\statesp}{state space}

\newcommand{\transp}[1]{{#1}{}^\top}

\newcommand{\obser}{\ensuremath{a}}     

\newcommand{\pde}{\partial}

\renewcommand{\det}{\mbox{\rm det}\,}

\newcommand{\msr}{\ensuremath{\rho}}                

\newcommand{\BER}[1]{{\mbox{\footnotesize BER}}} 

\newcommand{\pS}{\ensuremath{{\cal M}}}          
\newcommand{\ssp}{\ensuremath{x}}                
\newcommand{\Lop}{\ensuremath{{\cal L}}}       

\newcommand\map{f}                  

\newcommand\xInit{{x_0}}        

\newcommand{\cl}[1]{{\ensuremath{n_{#1}}}}   

\newcommand{\monodromy}{\ensuremath{M}}   

\newcommand{\ExpaEig}{\ensuremath{\Lambda}}

\newcommand{\eigExp}[1][]{
     \ifthenelse{\equal{#1}{}}{\ensuremath{\lambda}}{\ensuremath{\lambda^{(#1)}}}}
\newcommand{\eigRe}[1][]{
     \ifthenelse{\equal{#1}{}}{\ensuremath{\mu}}{\ensuremath{\mu^{(#1)}}}}
\newcommand{\eigIm}[1][]{
     \ifthenelse{\equal{#1}{}}{\ensuremath{\omega}}{\ensuremath{\omega^{(#1)}}}}

\newcommand{\nws}{non--wandering set}

\newcommand{\diffTen}{\ensuremath{\Delta}}  
\newcommand{\covMat}{\ensuremath{Q}}             
\newcommand{\Lnoise}[1]{{\cal L}_{FP}^{#1}} 
\newcommand{\orbitDist}{\ensuremath{z}}     

\newcommand {\id}{{\ \hbox{{\rm 1}\kern-.6em\hbox{\rm 1}}}}

\newdimen\onebox
\newdimen\boxsize
\newcount\boxnum
\gdef\mult#1#2#3{
    \ifx#1\relax\else%
      \ifx#2\relax\else%
        #1=#2%
        \ifx#3\relax\else%
          \multiply#1#3%
        \fi%
      \fi%
    \fi}

\newlength{\verti}

\thicklines
\newlength{\Fsize}   
\newlength{\Fdotsize}
\ifboyscout 
  \newcommand{\PC}[1]{$\footnotemark\footnotetext{Predrag: #1}$}
  
  \newcommand{\JG}[1]{$\footnotemark\footnotetext{Gibson: #1}$}
  
  \newcommand{\JMH}[1]{$\footnotemark\footnotetext{Jeffrey: #1}$}
  
  \newcommand{\ES}[1]{$\footnotemark\footnotetext{Siminos: #1}$}
  
  \newcommand{\DL}[1]{$\footnotemark\footnotetext{Domenico: #1}$}
  
  \newcommand{\GMW}[1]{$\footnotemark\footnotetext{Gable: #1}$}
  
  \newcommand{\DK}[1]{$\footnotemark\footnotetext{Daniel: #1}$}
  
  \newcommand{\HF}[1]{$\footnotemark\footnotetext{Fogedby: #1}$}
\else 
  \newcommand{\PC}[1]{}
  
  \newcommand{\JG}[1]{}
  
  \newcommand{\JMH}[1]{}
  
  \newcommand{\ES}[1]{}
  
  \newcommand{\DL}[1]{}
  
  \newcommand{\GMW}[1]{}
  
  \newcommand{\DK}[1]{}
  
  \newcommand{\HF}[1]{}
\fi  

\begin{document}

\title{Mapping densities in a noisy state space}

\author{Domenico Lippolis}

\address{
Department of Physics, Pusan National University\\
Busan 609-735, South Korea \\
}

\maketitle

\abstract
Weak noise smooths out fractals in a chaotic \statesp\ and
introduces a maximum attainable resolution to its structure. The
balance of noise and deterministic stretching/contraction in
each neighborhood introduces local invariants of the dynamics
that can be used to partition the \statesp. We study the local
discrete-time evolution of a density in a two-dimensional
hyperbolic \statesp, and use the asymptotic eigenfunctions for
the noisy dynamics to formulate a new \statesp\ partition
algorithm.
\endabstract
\section{Motivation and outline}
Chaotic systems' main feature is their high sensitivity to
initial conditions. That makes direct numerical integration of
the equations difficult and often calls for alternative methods
for the evaluation of long-time averages of observables, such
as decay of correlations, diffusion coefficients, energy spectra,
or escape rates~\rf{ruelle}. To properly weigh these averages,
one needs to understand which regions of the \statesp\ are more
or less relevant for the dynamics, in other words make a
\textit{partition}~\rf{Kitch98}. Invariants of the dynamics such
as unstable periodic oribts have been successfully used to
partition the \statesp~\rf{DasBuch}. 

However, noise, modelled by
stochastic variables, erases periodic orbits. One has to look for
new invariants. For that reason, we previously have
studied~\rf{LipCvi08,CviLip12} the evolution of \textit{densities}
of trajectories and determined eigenfunctions of the local Fokker-Planck
operator in the vicinity
of the deterministic periodic orbits. The eigenfunctions are then
used to partition the \statesp. All that was done in discrete time in one
dimension. In order to develop a similar algorithm in higher
dimensions, the first step is again to study the evolution of
densities in the neighborhood
of the periodic points of the deterministic system. 

In the
present contribution we focus on the asymptotic evolution in
two dimensions, forward and backward in time of a
noiseless hyperbolic map (\refsect{det_ev}), to which we successively add weak, uncorrelated,
isotropic noise (\refsect{isot_ns}). In both cases
the densities asymptotically align with the unstable (stable) direction of the monodromy
matrix when iterated forward (backward) in time.
We finally use our results to propose 
a definition of a \textit{neighborhood} for an optimal
partition of the \statesp.

\section{Deterministic evolution}
\label{det_ev}
We start by reviewing the deterministic evolution of densities and observables
in the neighborhood of a fixed point $\mathbf{x}_0$ of the two-dimensional map
$\mathbf{x}'=\mathbf{f(x)}$. We assume that the fixed point is
hyperbolic, \ie,  that the Jacobian matrix
evaluated at the fixed point,
\beq
\monodromy_{ij}(\mathbf{x}_0)
  =  \left. \frac{ \pde \map_i(\mathbf{x})}{\pde \ssp_j} \right|_{\mathbf{\ssp}=\mathbf{\xInit}}
\,, \label{jac_0}
\ee{DL:JacobianMatrix}
has  eigenvalues $|\ExpaEig_s|<1$, $|\ExpaEig_u|>1$.

Consider the simplest example, a map
\beq
\mathbf{f(x)} = \left(\ExpaEig_s x,\ExpaEig_u y\right)
\ee{DL:rugh_map}
that is contracting along the $x$-axis and expanding along the $y$-axis.
Now consider a density of trajectories $\rho(\mathbf{x})$, for instance a Gaussian placed around the fixed point of $\mathbf{f(x)}$, and apply the Perron-Frobenius
operator~\rf{DasBuch} to it:
\beq
\Lop \,\rho(\mathbf{x}) = \int d\mathbf{z}\, \delta\left(\mathbf{x}-\mathbf{f(z)}\right)
\rho(\mathbf{z})
= \frac{1}{|\ExpaEig_u\ExpaEig_s|}\rho\left(\frac{x}{\ExpaEig_s},\frac{y}{\ExpaEig_u}\right)
\,,
\ee{DL:one_step}
so that, after $n$ iterations,
\beq
\Lop^\cl{} \rho(\mathbf{x}) = \frac{1}{|\ExpaEig_u|^\cl{}}\frac{\rho(\frac{x}{\ExpaEig_s^\cl{}},\frac{y}{\ExpaEig_u^\cl{}})}{|\ExpaEig_s|^\cl{}}
\,.
\ee{DL:n_steps}
One can see this as a density, which is losing mass by a factor of
$|\ExpaEig_u|^{-1}$ at each iteration.
This expression can be renormalized by a
factor of $|\ExpaEig_u|^\cl{}$, when taking the limit $n\rightarrow\infty$.
If the initial density is a normalized Gaussian
$\rho(\mathbf{x}) \propto \exp\left[-(x^2+y^2)/2\sigma^2\right]$, we obtain
\beq
\lim_{\cl{}\to\infty} |\ExpaEig_u|^\cl{} \Lop = \lim_{\cl{}\to\infty}
\frac{\exp\left[-\frac{x^2}{2(\sigma\ExpaEig_s^\cl{})^2}\right]}{\sqrt{2\pi\sigma^2\ExpaEig_s^{2\cl{}}}}
= \delta(x)
\ee{DL:lim_dens}
meaning the limiting density is supported on the $y$-axis, the \textit{unstable manifold} of the fixed point of the map.

Of course this was the simplest possible example, given that the contracting and expanding directions of the fixed point are already separated by the coordinates.
This is not the case in general, and one needs to do something different from what we just described.
We will follow
Rugh's formalism\rf{hhrugh92} for a general
two-dimensional map $\mathbf{f(x)}$ with a hyperbolic fixed point  $\mathbf{x}_0$:
the equation $f_y(x_i,y_i) = y_f$ has a unique solution, which we can call
$\phi_s(x_i,y_f)$, which is analytic and a contraction.
On the other hand, one can define
\beq
\phi_u(x_i,y_f) = f_x\left(x_i,\phi_s(x_i,y_f)\right)
\ee{DL:rugh_phiu}
and then rewrite $\mathbf{f}$,
\beq
\mathbf{f}\left(x_i,\phi_s(x_i,y_f)\right) = \left(\phi_u(x_i,y_f),y_f\right)
\,,
\ee{DL:isomorphism}
in terms of the \textit{pinning coordinates} $(x_i,y_f)$, that is the contracting coordinate of the initial point, $x_i$ and the expanding coordinate of the final point, $y_f$.
It is important to remark that both $\phi_u(x_i,y_f)$ and $\phi_s(x_i,y_f)$ are contractions
on their supports\rf{hhrugh92} . In particular, for fixed $x_i$,
\beq
\lim_{n\rightarrow\infty}\phi_s^\cl{}(x_i,y_f) = W^s(x_i)
\ee{DL:stab_man}
with  $W^s(x_i)$ such that $\left(x_i,W^s(x_i)\right)$ parametrizes the stable manifold of the map
$\mathbf{f(x)}$. Similarly,
 \beq
\lim_{n\rightarrow\infty}\phi_u^\cl{}(x_i,y_f) = W^u(y_f)
\ee{DL:unstab_man}
where $\left(W^u(y_f),y_f\right)$ defines the unstable manifold of $\mathbf{f(x)}$.

Now we will obtain again the limit \refeq{DL:lim_dens} for the evolution of a density
carried by the Perron-Frobenius operator.
We will study the evolution \refeq{DL:n_steps}
inside a space average of an observable $\obser(\mathbf{x})$.
\beq
\left<a\right>_n = \frac{1}{|{\cal{M}}|}\int_{{\cal{M}}}d\mathbf{x}\,\obser(\mathbf{x})\left[\Lop^\cl{}
\rho(\mathbf{x})\right]
\ee{DL:aver_a}
As we will show, the support of the observable inside the average corresponds to the support of the
mapped density.
Eq.~\refeq{DL:aver_a} is equivalent to a more familiar expression for the space average,
which we can easily write by letting $\Lop^\cl{}$ act on its left on the
observable $\obser$, in which case it becomes the Koopman operator ${\cal{K}}$\rf{DasBuch},
\begin{eqnarray}
\nonumber
\int_{{\cal{M}}}d\mathbf{x}\obser(\mathbf{x})\left[\Lop^\cl{}\rho(\mathbf{x})\right] &=&
\int_{{\cal{M}}}d\mathbf{x}\left[{\cal{K}}^\cl{}\obser(\mathbf{x})\right]\rho(\mathbf{x}) \\
&=&
\int_{{\cal{M}}}d\mathbf{x}\,\obser\left(\mathbf{f}^\cl{}(\mathbf{x})\right)\rho(\mathbf{x})
\label{DL:koop}
\end{eqnarray}
We now change the coordinates in the last integral according to the transformation
\refeq{DL:isomorphism}:
\beq
 \int_Idx_idy_fa\left(\phi^\cl{}_u(x_i,y_f), y_f\right)\rho\left(x_i, \phi_s^\cl{}(x_i,y_f)\right)
\det\left(\partial_2\phi_s^\cl{}(x_i,y_f)\right)
\ee{DL:transf_aver}
where $I$ is the new domain of integration and $\det\left(\partial_2\phi_s^\cl{}(x_i,y_f)\right)$ is the
determinant of the Jacobian of the change of coordinates in the integral.
For $n\rightarrow\infty$ one gets, according to \refeq{DL:stab_man} and \refeq{DL:unstab_man},
\beq
 \int_Idx_idy_fa\left(W^u(y_f), y_f\right)\rho\left(x_i, W^s(x_i)\right)
\det\left(\partial_2\phi_s^\infty(x_i,y_f)\right).
\ee{DL:transf_aver1}
We can see that the observable $\obser$ ends up being supported on the \textit{unstable manifold} of the map.
Intuitively, the initial observable
stretches and contracts respectively along the unstable and stable manifolds, so that it asymptotically survives on the only region of the \statesp\ (the unstable manifold) where it cannot be crushed by the contraction.
In the separable case \refeq{DL:rugh_map}, the average~\refeq{DL:aver_a}
is proportional to
\bea
\nonumber
&\int dxdy\, \obser(x,y)\Lop^\cl{}\rho(x,y) \rightarrow \frac{1}{|\ExpaEig_u|^n}\int dxdy\, \obser(x,y)\delta(x) = \\
&\frac{1}{|\ExpaEig_u|^n}\int dy\, \obser(0,y).
\label{trv_crscheck}
\eea
In other words, knowing the support of the observable $\obser$ inside the average
is equivalent to knowing the support of the time-forward evolved density $\rho$
(in this case the $y-$axis, cf. \refeq{DL:lim_dens}).

Now we consider the time-backward evolution, described by the
Koopman operator ${\cal{K}}$:
\bea
\nonumber
&\int_{{\cal{M}}}d\mathbf{x}\,\obser(\mathbf{x})
     \left[{\cal{K}}^\cl{}\rho(\mathbf{x})\right] \rightarrow \\
&\int_Idx_idy_f\,\obser\left(x_i, W^s(x_i)\right)\rho\left(W^u(y_f), y_f\right)
\det\left(\partial_2\phi_s^\infty(x_i,y_f)\right)
\label{DL:koop_on_right}
\eea
where we switched to pinning coordinates as in
\refeq{DL:transf_aver1}, with $\rho$ and $\obser$ inverted with respect to the
previous case, and then took the limit $n\rightarrow\infty$. This
time the observable $\obser$ is asymptotically supported on the \textit{stable manifold}.

Analogous results are found
for a periodic point $\mathbf{x}_a$ of a map $\mathbf{f(x)}$, for it can be
regarded as the fixed point of the iterated map
$\mathbf{f}^{\cl{p}}(\mathbf{x})$, with $\cl{p}$ period of the cycle.
The time-forward (backward) evolution aligns observables and thus densities
to the unstable (stable) eigenvector of the \textit{monodromy} matrix
\beq
\monodromy^{\cl{p}}_{ij}(\mathbf{x}_a)
  =  \left. \frac{ \pde \map^{\cl{p}}_i(\mathbf{x})}{\pde \ssp_j} \right|_{\mathbf{x}=
\mathbf{x}_a}
\ee{DL:iter_jacob}
evaluated at the periodic point $\mathbf{x}_a$.

\section{Adding noise}
\label{isot_ns}

We now add weak
noise to the map $\mathbf{f}(\mathbf{x})$.
In the vicinity of any point $x_a$, Gaussian
densities are mapped forward in time by
the Fokker-Planck operator $\Lnoise{}$\rf{CviLip12,Risken96}
\bea
\msr_{a+1}(\orbitDist_{a+1})
    &=&
    \frac{1}{C_{a}}
\int [d\orbitDist_a] \,
e^{-\frac{1}{2}\transp{(\orbitDist_{a+1}- \monodromy_a \orbitDist_a)}
         {} \frac{1}{\Delta} {\,}
           (\orbitDist_{a+1}-\monodromy_a \orbitDist_a)
                -\transp{\orbitDist_a} {} \frac{1}{Q_a}{\,} \orbitDist_a}
    \continue
    &=&
     \frac{1}{C_{a+1}}
        \;
    e^{-\frac{1}{2}
      {\transp{\orbitDist_{a+1}}} {} \frac{1}{Q_{a+1}} {\,} \orbitDist_{a+1}}
\,,
\label{DL:stepLater}
\eea
where we defined local coordinates $z_a=x-x_a$.
Here the noise is described by the symmetric and positive definite
diffusion tensor $\diffTen$. If the density is a Gaussian distribution,
we can recast the problem in terms of its covariance matrix\rf{CviLip12}:
\beq
Q_{a+1}  =  \monodromy_{a} Q_{a} \transp{\monodromy_{a}}+\Delta
\,.
\label{AddVariances}
\eeq
The long-time limit is given by the fixed-point condition $\covMat_a=\covMat_{a+1}$, valid when the dynamics is
contracting ($\monodromy$ has all eigenvalues $|\ExpaEig_i|<1$).
This condition states that the covariance matrix must be
invariant under the combined action of the deterministic contraction and expansion by weak noise after one time step.
Let $S$ be the matrix which diagonalizes $\monodromy$.
If we make the further transformations
$\covMat \rightarrow S^{-1} \covMat\transp{\left(S^{-1}\right)} \equiv \hat{\covMat}$
and $\diffTen \rightarrow S^{-1} \diffTen\transp{\left(S^{-1}\right)} \equiv \hat{\diffTen}$, the solution to the
fixed-point condition for contracting maps reads~\rf{CviLip12}
\beq
\hat{\covMat}_{ij} = \frac{1}{1-\ExpaEig_i \ExpaEig_j} \hat{\diffTen}_{ij}.
\label{contr_sol}
\eeq

We obtain the time-backward evolution by taking the adjoint of the operator in~\refeq{DL:stepLater}. Like before,
an equation is derived for the mapping of the covariance matrix:
\beq
\monodromy_a \covMat_a \transp{\monodromy_a} = \covMat_{a+1} + \diffTen.
\ee{DL:FPexpand}
If the deterministic dynamics is expanding ($\monodromy$ has all eigenvalues $|\ExpaEig_i|>1$),
\refeq{DL:FPexpand} becomes a fixed point condition by setting $\covMat_a=\covMat_{a+1}$. We find that
\beq
\hat{\covMat}_{ij} = \frac{1}{\ExpaEig_i \ExpaEig_j -1} \hat{\diffTen}_{ij},
\label{exp_sol}
\eeq
where we applied the same diagonalization transformation by means of the matrix $S$.

\begin{figure}[tbp]
(a) \includegraphics[width=0.21\textwidth,angle=0]{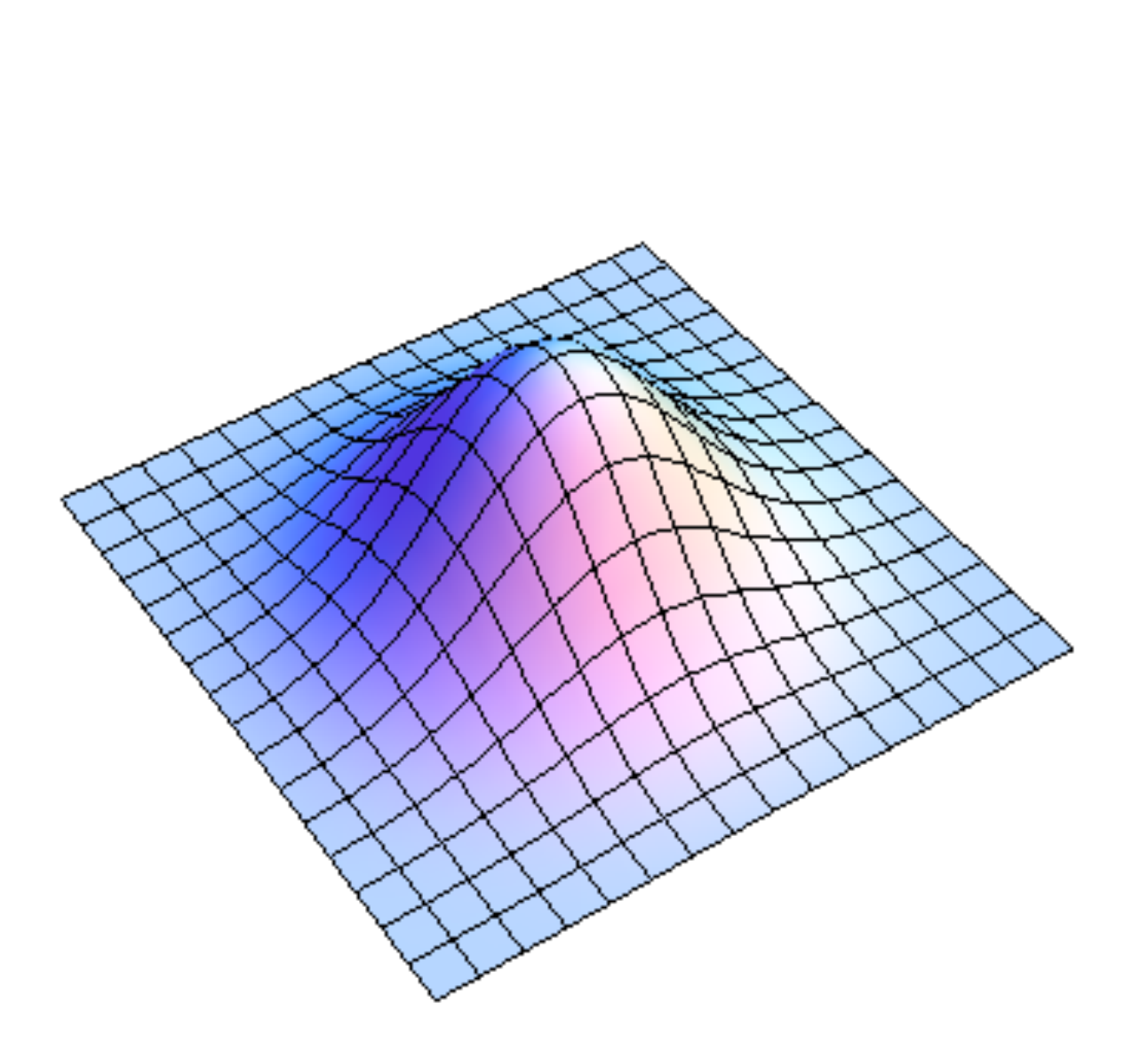}%
(b) \includegraphics[width=0.21\textwidth,angle=0]{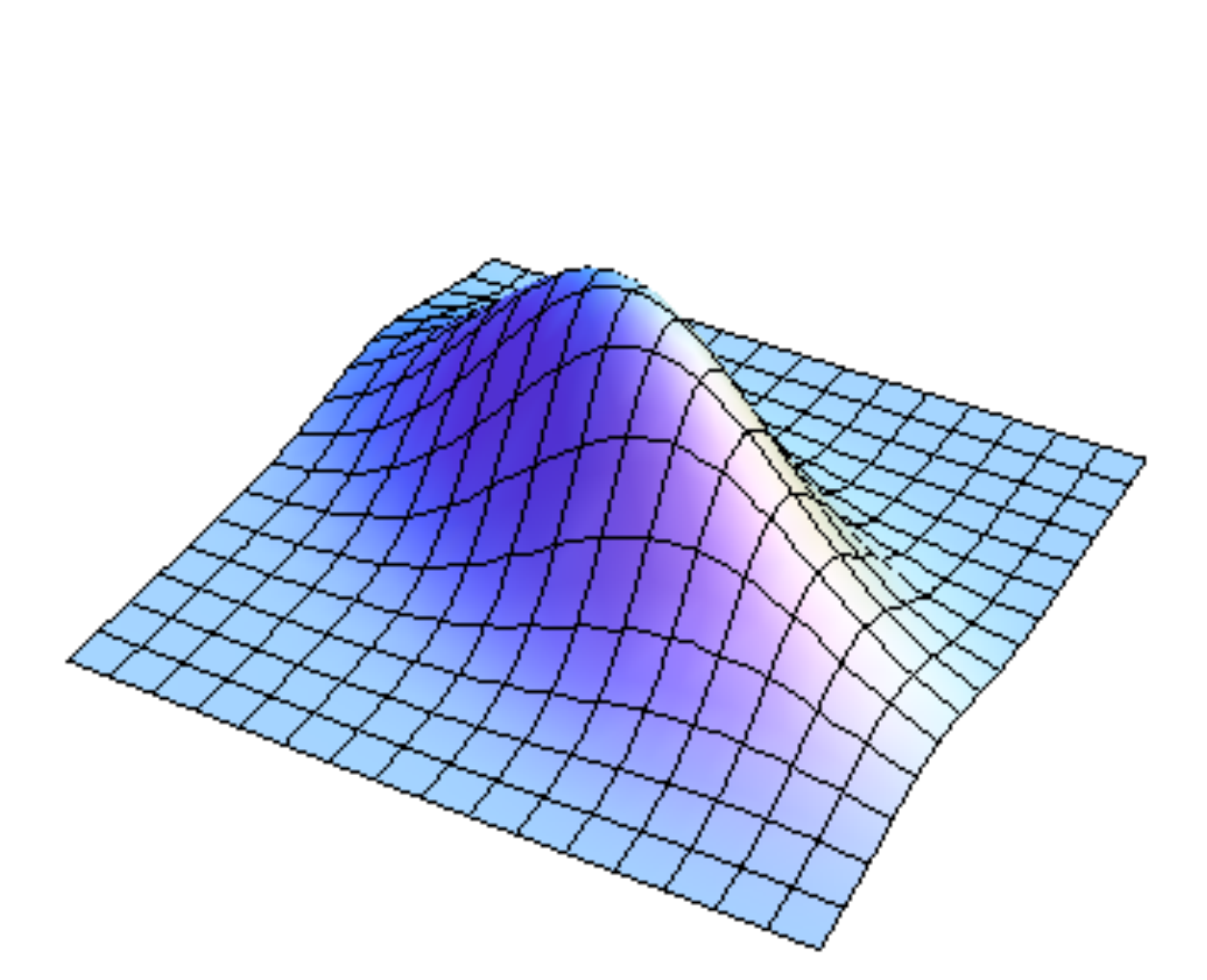}%
\\
(c) \includegraphics[width=0.21\textwidth]{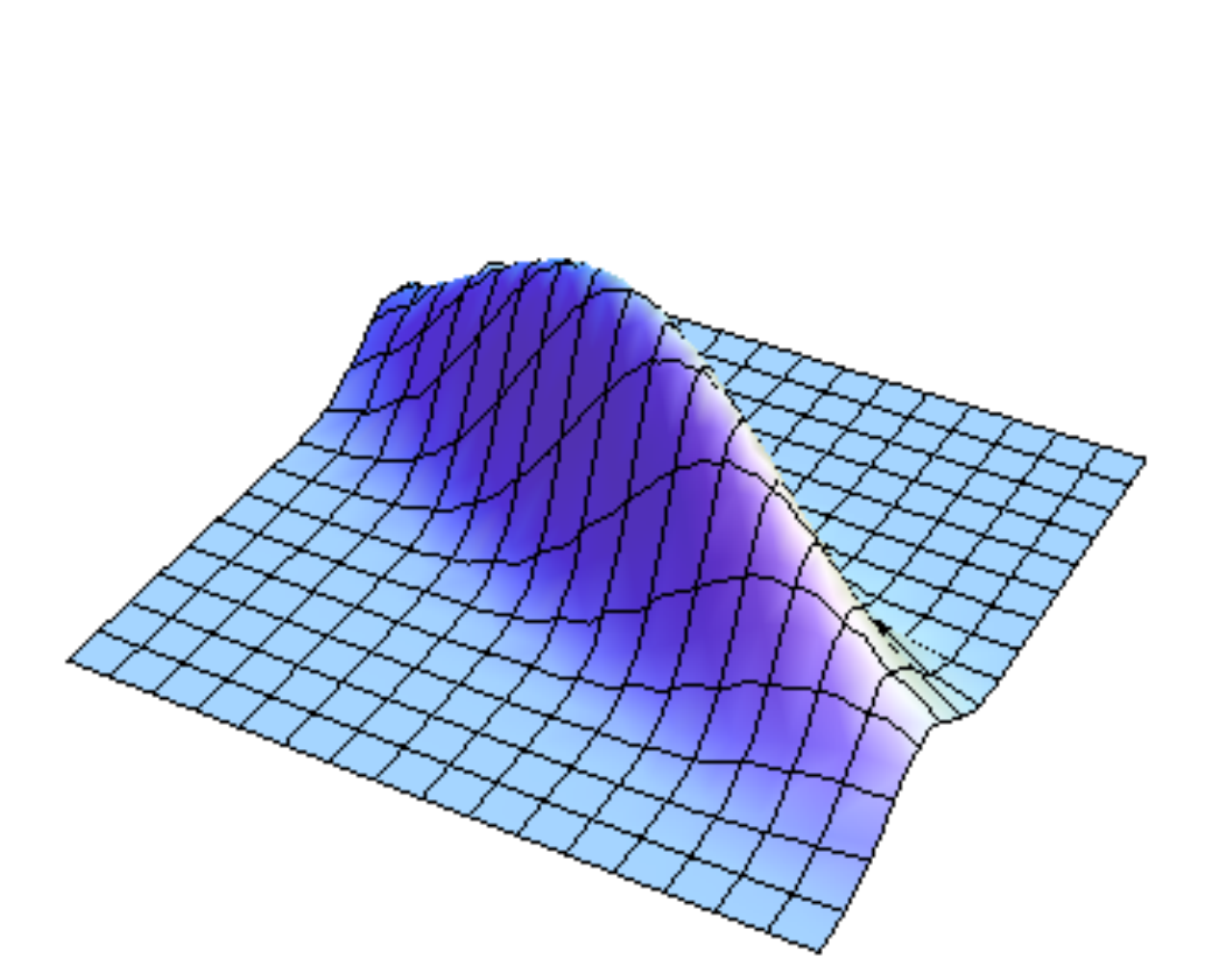}
(d) \includegraphics[width=0.21\textwidth]{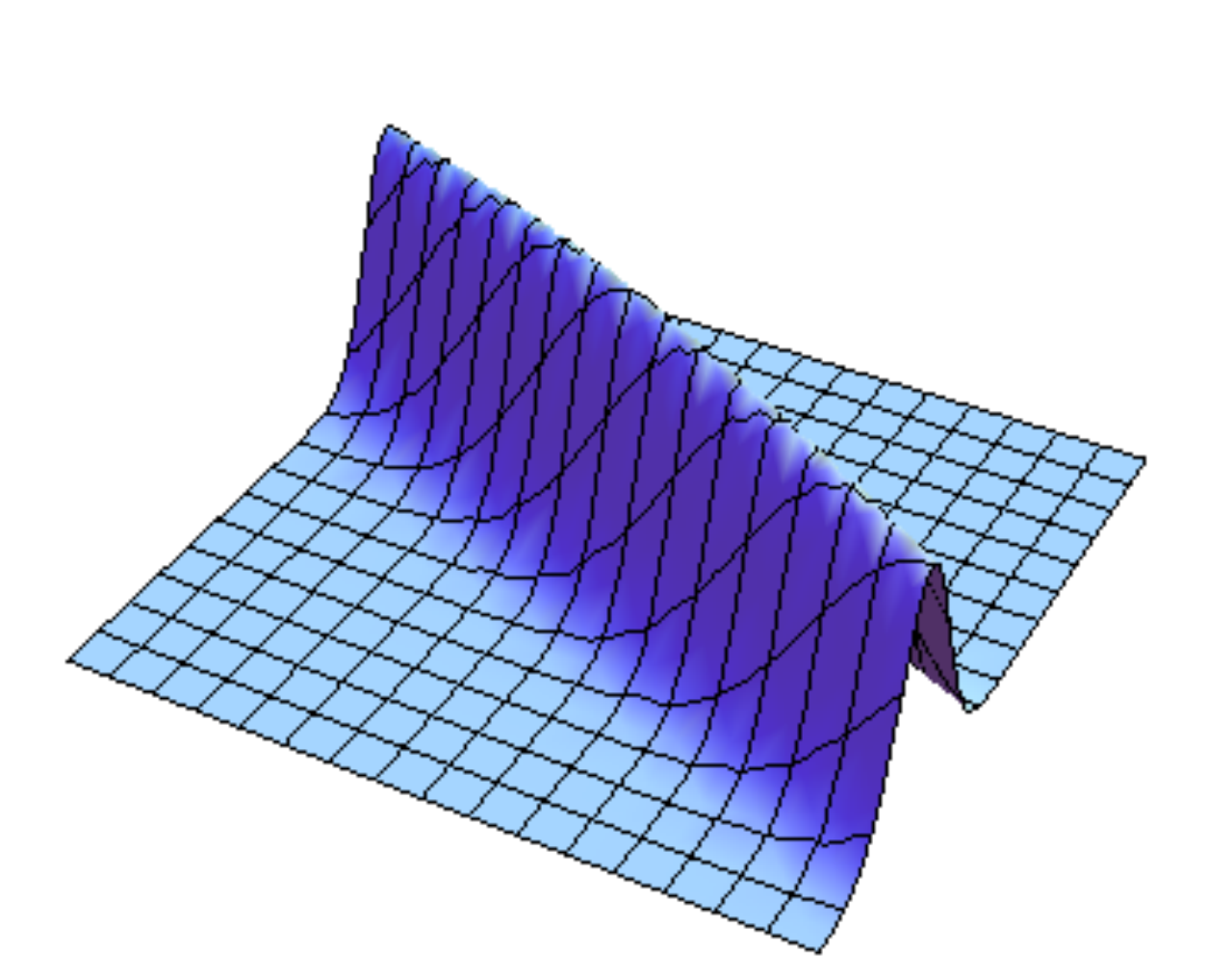}

\caption[]{
Evolution of a Gaussian density centered at the fixed point of a
hyperbolic map: (a) initial bump; (b) after one iteration; (c) after
two iterations; (d) asymptotic limit.}
\label{f:tubes}
\end{figure}

Typically in a chaotic system the matrix $\monodromy$ has both contracting and expanding directions,
so that neither the solution given by~\refeq{contr_sol} nor by~\refeq{exp_sol} applies.
In what follows we study the evolution of the covariant matrix both forward and
backward in time, looking for an asymptotic limit. We start forward in time, iterating~\refeq{AddVariances}
in the neighborhood of a fixed point of $\mathbf{f}(\mathbf{x})$:
\beq
Q_n = \Delta + \monodromy \Delta \transp{\monodromy}
        + \monodromy^2 \Delta \transp{(\monodromy^2)}
		+ \cdots +
            \monodromy^\cl{}Q_0\transp{(\monodromy^\cl{})}
\,.
\ee{ddQfixed}
As in the deterministic case, let us first see how the $Q$'s map when $\monodromy$ and
the initial $\covMat_0$ are diagonal: each matrix element $\covMat_{ii}$ obeys the sum \refeq{ddQfixed},
which diverges for $|\ExpaEig_i|>1$, so that $\covMat_{ii}^{-1}$ vanishes. On the other hand, it
converges to $\covMat_{ii}\rightarrow\Delta_{ii}/(1-\ExpaEig_i^2)$
\rf{LipCvi08} if $|\ExpaEig_i|<1$.
As a result, the axes of the Gaussian $e^{-\transp{z}Q^{-1}z}$
asymptotically survive in the stable directions only,
while the whole density is supported along the unstable directions, like in the deterministic
separable case (cf. \refeq{DL:lim_dens}).
In two dimensions, the asymptotic density looks like the Gaussian-shaped tube in \reffig{f:tubes}(d).

We will now explain the asymptotic evolution of the axes of the ellipsoid the Gaussian is
supported on when $\monodromy$ is not diagonal, mimicking an argument of Ott's\rf{ottbook}.
First of all, if the noise is isotropic, $\diffTen(\ssp) = 2\, D \, \mathbf{1}$,
the diffusion tensor $\diffTen$ is replaced by a scalar diffusion constant $D$ that can be
factored from
the sum~\refeq{ddQfixed}. Consider a vector $v$:
\beq
\monodromy^\cl{}\transp{(\monodromy^\cl{})} v =
\monodromy^\cl{}\sum_ja_j\ExpaEig_j^\cl{}\hat{e}_j'
\sim \monodromy^\cl{}a_u\ExpaEig_u^\cl{}\hat{e}_u'
\ee{mt_action}
where we first wrote $v$ in terms of the eigenvectors $\hat{e}_j'$ of
$\transp{\monodromy}$, then we applied $\transp{(\monodromy^\cl{})}$ to the same vector and
finally we observed that the result asymptotically aligns to the direction $\hat{e}_u'$ of the
most unstable eigenvalue of $\transp{\monodromy}$, $\ExpaEig_u$.
Let us now write $\monodromy^\cl{} = S\hat{\ExpaEig}^\cl{}S^{-1}$, with
$S  = (\hat{e}_1,...,\hat{e}_u,...,\hat{e}_N)$, and
$S^{-1} = \transp{(\hat{e}_1',...,\hat{e}_u',...,\hat{e}_N')}$.
So now
\begin{eqnarray}
\nonumber
\monodromy^\cl{}\ExpaEig_u^\cl{}\hat{e}_u' =
 S\hat{\ExpaEig}^\cl{}S^{-1}\ExpaEig_u^\cl{}\hat{e}_u' \propto
\\ \nonumber
S\hat{\ExpaEig}^\cl{}\transp{\left(
\hat{e}_u'\cdot\hat{e}_1',\dots,1,\dots, \hat{e}_u'\cdot\hat{e}_N'
                \right)}  =\\
S\transp{\left(
\ExpaEig_1^\cl{}\hat{e}_1'\cdot\hat{e}_u',\dots,\ExpaEig_u^\cl{},
\dots, \ExpaEig_N^\cl{} \hat{e}_N'\cdot\hat{e}_u'
\right)}
\label{mt_action2}
\end{eqnarray}
When $n\rightarrow\infty$ the $u$th component of the vector weighs above all others, making
the result proportional to
\beq
S \transp{\left(0,...,1,...,0\right)} = \hat{e}_u
\,.
\ee{mt_action3}
Thus any vector is eventually stretched and rotated toward
\textit{the most unstable direction} of
$\monodromy$.
It is straightforward to show that
the last term in the sum \refeq{ddQfixed},
$\monodromy^\cl{}Q_0\transp{(\monodromy^\cl{})}$, asymptotically behaves likewise.
Consequently, $Q_\infty$ and thus $Q_\infty^{-1}$ are also aligned
with the most unstable eigenvector of the monodromy matrix
$\monodromy$. Numerics (\reffig{f:tubes}) help us visualize the
result in two dimensions, where $\monodromy$ has one stable and one
unstable directions: an initial isotropic Gaussian develops into a
`tube', infinitely extended along the unstable manifold of
$\monodromy$, and having a Gaussian section in the orthogonal
direction, due to the balance of noise and deterministic contraction.

One can repeat the above reasoning when applying the adjoint Fokker-Planck operator
$\Lnoise{\dagger}$. In this case we invert~\refeq{DL:FPexpand} (the unknown being
$\covMat_a$), and iterate it $n$ times to get
\beq
\covMat_{-n} =\monodromy^{-n}\covMat_0\transp{(\monodromy^{-n})}
+\monodromy^{-n}\diffTen\transp{(\monodromy^{-n})} +
...+\monodromy^{-1}\diffTen\transp{(\monodromy^{-1})}
\,.
\ee{DL:ntimes_adj}
This is similar to~\refeq{ddQfixed}, except the monodromy matrix is
inverted, so that stable and unstable eigenvalues are swapped, and
the argument~\refeq{mt_action}-\refeq{mt_action3} results in any
vector being stretched and rotated toward \textit{the most stable
direction} of $\monodromy$.

The observations made for a fixed point of the map can be extended to a periodic orbit of
arbitrary period $n_p$.
We start again from \refeq{ddQfixed}:
\begin{eqnarray}
Q_{a} = \Delta + \monodromy_{a-1} \Delta \transp{(\monodromy_{a-1})}
+ \cdots +
\monodromy_{a-\cl{p}}^{\cl{p}}
    \covMat_{a-\cl{p}}\transp{(\monodromy_{a-\cl{p}}^{\cl{p}})}
\,.
\label{ddQvar}
\end{eqnarray}
We define
\bea
\Delta_{p,a} = \Delta
    +\monodromy_{a-1} \Delta \transp{(\monodromy_{a-1})}
    + \cdots
    +\monodromy_{a-\cl{p}+1}^{\cl{p}-1}
    \Delta \transp{(\monodromy_{a-\cl{p}+1}^{\cl{p}-1})},
\label{POnoiseHist}
\eea
and then the asymptotic evolution of the covariance matrix $Q$ around the periodic point $x_a$ can be
written as
\bea
Q_{p,a} =
         \monodromy_{p,a} \covMat_{a} \transp{\monodromy_{p,a}}
         +\Delta_{p,a}
\,,
\label{ddQpPoint}
\eea
where $\monodromy_{p,a}=\monodromy^{n_p}_a$, the latter defined in~\refeq{DL:iter_jacob}.
The problem reduces to the previous case, which
shows that the leading eigenvector of the asymptotic covariance matrix (or the major axis of the
ellipsoid) aligns to the most unstable (stable) eigenvector of $\monodromy_{p,a}$ in the
time-forward (backward) evolution.

\begin{figure}[tbp]
(a) \includegraphics[width=0.21\textwidth]{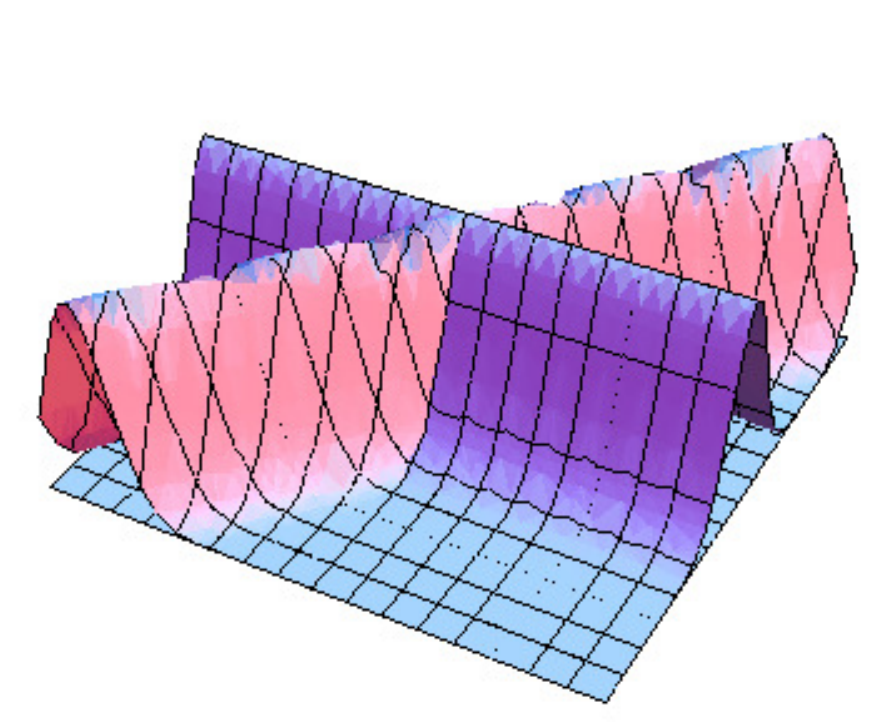}
(b) \includegraphics[width=0.21\textwidth]{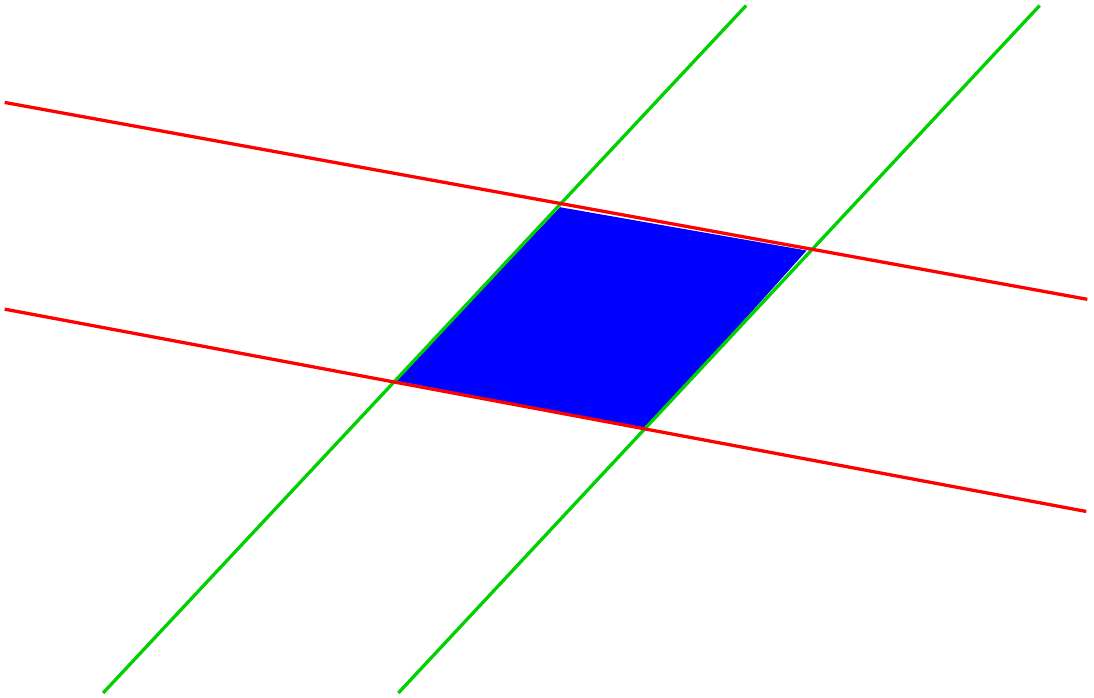}
\\
\centerline{
(c) \includegraphics[width=0.24\textwidth]{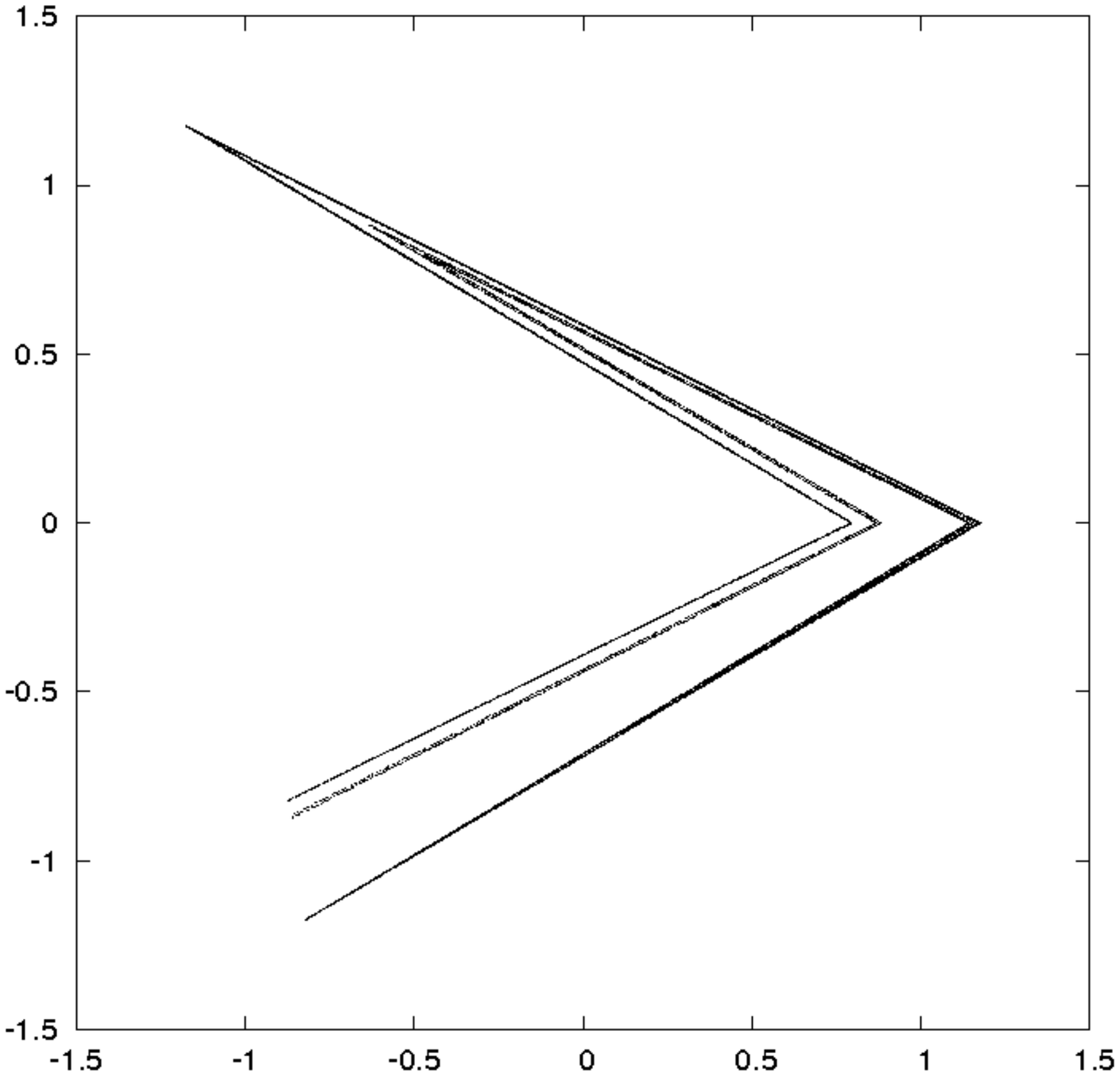}
(d) \includegraphics[width=0.21\textwidth]{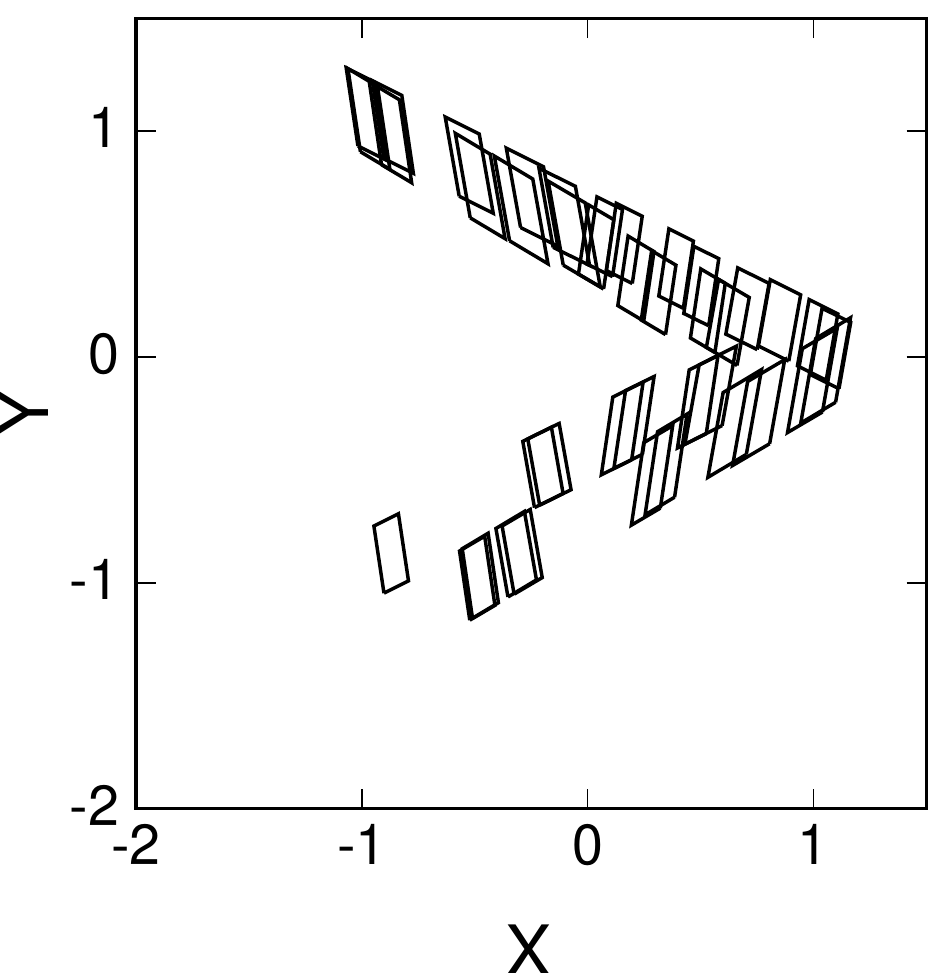}}
\caption{(a)
The ground-state eigenfunctions of $\Lnoise{}$ and of its adjoint
$\Lnoise{\dagger}$, both operator linearized around the same fixed point;
(b) our definition of partition interval in two dimensions: take the local
densities in (b), cut off their supports at $1\sigma$ and take their
intersections. The Lozi attractor~\rf{DasBuch}, (c) noiseless, and (d) noisy, covered
with neiborhoods from all periodic points up to length six.}
\label{f:neighbs}
\end{figure}

Following the technique we used in one dimension\rf{LipCvi08}, we
define the neighborhood $\pS_a$ of the periodic point $\ssp_a$ as the
intersection of the supports (within a 1$\sigma$ confidence) of the
ground-state local eigenfunctions of $\Lnoise{}$ and
$\Lnoise{\dagger}$. We use these regions to cover the \nws\ of the
system, starting with the periodic orbits of the shortest period, and
increasing the period until neighborhoods significantly overlap~(see
\reffig{f:neighbs}(d)).

\section{Summary}
We have studied the asymptotic evolution of Gaussian densities
of trajectories in the neighborhoods of hyperbolic periodic points,
first in a deterministic map and then in the presence of weak,
uncorrelated, isotropic noise. We investigated
both time-forward and -backward dynamics. The latter was
realized by means of the adjoint of the evolution operator.
As it turns out, the densities asymptotically align with the most unstable (stable) direction of the monodromy
matrix when iterated forward (backward) in time.
Using both asymptotic densities, we proposed a definition
for a \textit{neighborhood} which should then be used
to partition the \statesp.

\section*{Acknowledgments}
Special thanks to the NOLTA 2013 organizers for their invitation, as well as
to P. Cvitanovi\'c and H.H. Rugh for helpful correspondence. The author is 
also indebted to J.M. Heninger for carefully editing the manuscript.

\bibliographystyle{plain} 
\bibliography{../bibtex/lippolis}

\end{document}